\documentclass[12pt, preprint]{aastex}

\shorttitle{Masses and Radii for HAT-TR-205-013 }
\shortauthors{Beatty et al.}

\newcommand\kms{\ifmmode{\rm km\thinspace s^{-1}}\else km\thinspace s$^{-1}$\fi}
\newcommand\ms{\ifmmode{\rm m\thinspace s^{-1}}\else m\thinspace s$^{-1}$\fi}
\newcommand\msun{\ifmmode{{\cal M}_{\odot}}\else ${\cal M}_{\odot}$\fi}
\newcommand\rsun{\ifmmode{{\cal R}_{\odot}}\else ${\cal R}_{\odot}$\fi}
\newcommand\mjup{\ifmmode{{\cal M}_{\rm J}}\else ${\cal M}_{\rm J}$\fi}
\newcommand\rjup{\ifmmode{{\cal R}_{\rm J}}\else ${\cal R}_{\rm J}$\fi}

\begin{document}

\title{THE MASS AND RADIUS OF THE UNSEEN M-DWARF COMPANION IN THE
SINGLE-LINED ECLIPSING BINARY HAT-TR-205-013}

\author{\sc Thomas G. Beatty\altaffilmark{1},
        Jos\'{e} M. Fern\'{a}ndez\altaffilmark{2,3},
        David W. Latham\altaffilmark{2},
        G\'{a}sp\'{a}r \'{A}. Bakos\altaffilmark{2,4},
        G\'{e}za Kov\'{a}cs\altaffilmark{5},
        Robert W. Noyes\altaffilmark{2},
        Robert P. Stefanik\altaffilmark{2},
        Guillermo Torres\altaffilmark{2},
        Mark E. Everett\altaffilmark{6},
        Carl W. Hergenrother\altaffilmark{7,2}
}

\altaffiltext{1}{Department of Astronomy, Harvard University, 60
Garden Street, Cambridge, MA 02138}

\altaffiltext{2}{Harvard-Smithsonian Center for Astrophysics, 60
Garden Street, Cambridge, MA 02138}

\altaffiltext{3}{Department of Astronomy, Pontificia Universidad
Cat\'{o}lica, Casilla 306, Santiago 22, Chile}

\altaffiltext{4}{Hubble Fellow}

\altaffiltext{5}{Konkoly Observatory, Budapest, P.O. Box 67, H-1125, Hungary}

\altaffiltext{6}{Planetary Science Institute, 1700 East Fort Lowell
Road, Suite 106, Tucson, AZ 85719}

\altaffiltext{7}{Lunar and Planetary Laboratory, University of
Arizona, Tucson, AZ 85719}

\email{dlatham@cfa.harvard.edu}

\begin{abstract}
We derive masses and radii for both components in the single-lined eclipsing binary HAT-TR-205-013, which consists of a F7V primary and a late M-dwarf secondary. The system's period is short, $P=2.230736 \pm 0.000010$ days, with an orbit indistinguishable from circular, $e=0.012 \pm 0.021$. We demonstrate generally that the surface gravity of the secondary star in a single-lined binary undergoing total eclipses can be derived from characteristics of the light curve and spectroscopic orbit. This constrains the secondary to a unique line in the mass-radius diagram with $M/R^2$ = constant. For HAT-TR-205-013, we assume the orbit has been tidally circularized, and that the primary's rotation has been synchronized and aligned with the orbital axis. Our observed line broadening, $V_{\rm rot} \sin i_{\rm rot} = 28.9 \pm 1.0$ \kms, gives a primary radius of $R_{\rm A} = 1.28 \pm 0.04$ \rsun. Our light curve analysis leads to the radius of the secondary, $R_{\rm B} = 0.167 \pm 0.006$ \rsun, and the semimajor axis of the orbit, $a = 7.54 \pm 0.30\ \rsun\ = 0.0351 \pm 0.0014$ AU. Our single-lined spectroscopic orbit and the semimajor axis then yield the individual masses, $M_{\rm B} = 0.124 \pm 0.010$ \msun\ and $M_{\rm A} = 1.04 \pm 0.13$ \msun. Our result for HAT-TR-205-013 B lies above the theoretical mass-radius models from the Lyon group, consistent with results from double-lined eclipsing binaries. The method we describe offers the opportunity to study the very low end of the stellar mass-radius relation.
\end{abstract}

\keywords{binaries: eclipsing --- binaries: spectroscopic --- stars:
fundamental parameters --- stars: low-mass --- stars: rotation}

\section{INTRODUCTION}

Solving for the masses and radii of stars has traditionally been
accomplished through the analysis of double-lined eclipsing binaries,
where the light of both components is detected. Masses and radii
determined this way are fundamental and can be very accurate, because
they rely only on Newton's laws and geometry for the analysis of the
spectroscopic orbit and light curve, and not on models of stellar
structure and evolution. In particular, analysis of the eclipse light
curve yields the orbital inclination, and when combined with the
double-lined spectroscopic orbit, this yields individual masses for
both stars.

There are dozens of double-lined eclipsing binaries with very
accurate mass and radius determinations \citep[e.g. see][for a
review]{andersen1991}, but only 10 M dwarfs (in 5 systems) with
accuracies better than 3 percent. In order of increasing mass, the 5
systems are: CM Draconis \citep{lacy1977, metcalfe1996}, CU Cancri
\citep{ribas2003}, NSVS01031772 \citep{lopez2006}, YY Geminorum
\citep{torres2002}, and GU Bo\"{o}tis \citep{lopez2005}.  Figure
\ref{fig1} shows the observational results for these 10 M dwarfs on
the mass-radius diagram, along with the predicted theoretical models
from \cite{baraffe1998}. All of the observed radii are larger than the
theoretical predictions, typically by 5 to 10 percent. Another
striking feature of Figure \ref{fig1} is the lack of accurate
mass-radius determinations in the lower half of the M-dwarf mass
range, from CM Dra B (0.214 \msun), all the way down to the substellar
limit near 0.075 \msun. Recently, however, the growing number of
short-period single-lined eclipsing binaries with F- and G-star
primaries and M-dwarf secondaries identified by wide-angle photometric
surveys for transiting planet promises to provide a way to fill in this
gap in the mass-radius diagram \citep{bouchy2005, pont2005a,
pont2005b, pont2006}.

One approach to using a single-lined eclipsing binary to solve for the
mass and radius of the unseen companion is to use stellar models
together with spectroscopic and photometric observations of the
primary to estimate its mass and radius. Then the radius ratio from
the light curve yields the radius of the secondary, and the mass
function from the spectroscopic orbit with the orbital inclination
from the light curve can be combined to yield the mass of the
secondary. However, this approach is not fundamental, because it
relies on stellar models to characterize the primary. Tests of the
mass-radius relation with the low-mass secondaries in such systems are
no better than the validity of the models for the primaries. In
particular, the theoretical isochrones depend on metallicity, and
significant errors can result if the wrong metallicity is adopted.

An alternative, and more fundamental, approach relies on the
expectation that short-period binaries whose orbits have been
circularized by tidal mechanisms must also have the axial rotation of
both stars synchronized to the orbital period, and both rotational
axes aligned with the normal to the orbit, so that the two
inclinations are equal, $i_{\rm rot} = i_{\rm orb}$ \citep[e.g.\
see][]{Hut81,Zahn89}. In this case, a measurement of the spectral
line broadening due to rotation, $V_{\rm rot} \sin i_{\rm rot}$,
combined with a value for $i_{\rm orb}$ from an analysis of the light
curve, yields the actual equatorial rotational velocity, $V_{\rm
rot}$. If the rotation is synchronized with the orbit, then $P_{\rm
rot} = P_{\rm orb}$, and the radius of the primary can be solved. The
key to this approach is the ability to derive accurate values for the
rotational broadening of the spectral lines, because the primary
radius can be no more accurate than the value derived for $V_{\rm
rot}$. The radius of the primary then sets the scale for the rest of
the system, yielding the radius of the secondary and the semi-major
axis of the orbit in actual length units, as well as the orbital
inclination. Then Newton's revised version of Kepler's Third Law can
be used to derive the sum of the masses, and the mass function from
the single-lined spectroscopic orbit allows the individual masses to
be determined. This approach depends on the predictions from stellar
models only in minor ways: limb darkening coefficients are needed for
the detailed analysis of the eclipse light curve, and the rotational
broadening that is derived from the observed spectra can depend weakly
on the metallicity that is adopted.

The advantage of using single-lined eclipsing binaries is that it
dramatically increases the number of low-mass stars whose masses and
radii can be determined. Indeed, over the past 30 years only seven
double-lined eclipsing systems composed of low-mass stars have been
identified (the aforementioned CM Dra, CU Cnc, NSVS01031772, YY Gem,
and GU Boo, plus OGLE-BW3-V38 \citep{maceroni2004}, and
TrES-Her0-07621 \citep{creevey2005}). Meanwhile, more than 75
single-lined eclipsing binaries with M-dwarf secondaries have been
discovered by wide-angle photometric transiting planet surveys such as Vulcan, TrES, and HAT in just the last five
years \citep{Latham2007}.

Increasing the number of low-mass stars with fundamental
determinations of masses and radii is worthwhile because of the
insights these systems can yield into stellar structure.  Low-mass
stars near the hydrogen-burning limit are cool enough that their
interior temperatures are on the order of the electron Fermi
temperature \citep{chabrier1997}, causing parts of the stellar
interior to be in the state of a partially degenerate electron gas,
which means a classical Maxwell-Boltzmann description of the interior
does not apply. Furthermore, the electron number density is such that
the mean inter-ionic distance is itself on the order of the
Thomas-Fermi screening length, meaning that the electron gas is
polarized by the external-ionic field \citep{chabrier1997}. Add to
this the further complexity that magnetic fields may play a role in
the inner workings of low mass stars \citep{mullan2001}, and it
becomes clear that any attempt to describe the interior of low mass
stars must take into account the physics of both magnetic fields and
partially degenerate polarized plasmas.

Not only is it difficult to model the interiors of low-mass stars, but
the usual grey model atmospheres are no longer applicable. The
relatively low temperatures of the stellar atmospheres allow for the
recombination of molecular hydrogen and other molecules, such as TiO.
Therefore, accurate non-grey model atmospheres that take into account the
effects of molecules must be derived and matched to the interior
models \citep{baraffe2002}.

Thus stars near the bottom of the main sequence pose a challenge to
dense matter physicists and stellar astronomers. One of the few
methods of testing models for low-mass stars is by confronting
theoretical predictions of the mass-radius relation with observations.
Up until now there has been little data with which to constrain the
possible theories. Measuring the masses and radii of M-dwarfs in
single-lined eclipsing binaries therefore provides a new opportunity to test the
mass-radius relation near the bottom of the main sequence.

In this paper, we determine masses and radii for the components of
HAT-TR-205-013, an eclipsing single-lined binary identified by the
HATnet (Hungarian-made Automated Telescope Network, see \cite{bakos2004}). The
unseen M-dwarf secondary has a mass and radius of 0.124 \msun\ and
0.167 \rsun, with 1-$\sigma$ errors of 9 and 4 percent, respectively. This
result places the M dwarf about 10 percent above the radius predicted
by the \citet{baraffe1998} models. In this first paper we describe
our analysis techniques in detail. In future papers we will present
the results for additional M dwarf secondaries.

\section{OBSERVATIONS AND DATA REDUCTION}

\subsection{HAT Photometry}

The HATnet project,\footnote{www.hatnet.hu} initiated in 2003 by
G.~\'A.~B, is a wide-field survey that aims for the discovery of
transiting planets around bright stars. It currently comprises 6 small
wide-field automated telescopes, each of which monitors
$8\arcdeg\times8\arcdeg$ of sky, typically containing 5000 stars
bright enough to permit detection of planetary transits via the
typical 1\% photometric dips they induce on their parent stars. The
instruments are deployed in a two-station, longitude-distributed
network, with four telescopes at the Fred L.\ Whipple Observatory in
Arizona, and two telescopes at the Submillimeter Array at Hawaii. For
a more detailed description of HAT's instrumentation, observations,
and data flow, see \cite{bakos2002,bakos2004}.

The HAT-TR-205-013 system lies in HATnet survey field G205, centered at $\alpha = 22^h55^m$, and $\delta = +37\arcdeg30\arcmin$. 3357 observations were made of this field by the HATnet telescopes between 5 October 2003 and 30 January 2004. Exposure times were 5 min at a cadence of 5.5 min. Light curves were derived by aperture photometry for the 6400 stars in G205 bright enough to yield photometric precision of better than 2\% (reaching 0.3\% in some cases). In deriving the light curves, we made use of the Trend Filtering Algorithm \citep{kovacs05} to correct for spurious trends in the data. We then searched all the light curves for characteristic transit signals, using the Box-fitting Least Squares \citep{kovacs02} algorithm, which searches for box-shaped dips in the parameter space of frequency, transit duration, and phase of ingress. Candidate transit signals with the highest detection significance were then examined individually, to isolate those with the best combination of stellar type (preferably main-sequence stars of spectral type mid-F or later), and depth, shape, and duration of transit.  One of these was HAT-TR-205-013, for which we identified a prominent periodicity with period 2.2307 days and transit depth of 0.02 mag. Figure \ref{fig2} shows the phase-folded and flux-normalized data for HAT-TR-205-013.

Following our determination of HAT-TR-205-013 as a potential planetary
system, we identified the star in the 2MASS catalog as 2MASS
23080834+3338039, which yielded $J$ and $K$ magnitudes of $J=9.691$ and
$K=9.408$. We also found HAT-TR-205-013 in the Tycho catalog, as TYC
2755-36-1, with magnitudes $B_{\rm T} = 11.355$ and $V_{\rm T} = 10.729$.
We then scheduled HAT-TR-205-013 for follow-up spectroscopic
observations to determine if the photometric dip was caused by a
stellar companion.

\subsection{Follow-up Spectroscopy} 

Our usual strategy for following up transiting-planet candidates
identified by wide-field photometric surveys is to start with an
initial spectroscopic reconnaissance, to see if there is evidence for
a stellar companion that is responsible for the observed light curve.
We have used the CfA Digital Speedometers \citep{latham1992} on the
1.5-m Wyeth Reflector at the Oak Ridge Observatory in the Town of
Harvard, Massachusetts and on the 1.5-m Tillinghast Reflector at the
Fred L.\ Whipple Observatory on Mount Hopkins, Arizona to obtain
single-order echelle spectra in a wavelength window of 45 \AA\
centered at 5187 \AA\, with a resolution of 8.5 \kms\ and a typical
signal-to-noise ratio per resolution element of 15 to 20.  For slowly
rotating solar-type stars these spectra deliver radial velocities
accurate to about 0.5 \kms, which is sufficient to detect orbital
motion due to companions with masses down to about 5 or 10 \mjup\ for
orbital periods of a few days.

Spectroscopic follow-up observations of transiting-planet candidates
are easier to schedule than photometric observations, because the
radial velocity varies throughout the entire orbital phase, while the
transit light curve has a duty cycle of only a few percent, e.g.\ 3
hours out of 3 days. Spectroscopy has the additional advantage over
photometry that we can use our spectra to classify the parent star,
deriving effective temperature, rotational velocity, and surface
gravity by correlating the observed spectra with a library of
synthetic spectra. This information is often useful in rejecting host
stars that are too hot, or are rotating too rapidly, or are evolved
(in which case we assume the giant must be the bright star in a
blended triple). Thus at CfA we usually start with a spectroscopic
reconnaissance before attempting follow-up photometry or very precise
radial-velocity observations \citep[e.g.\ see][]{odonovan2007}.

Starting in 1999 with transiting-planet candidates provided by the
Vulcan team \citep{borucki2001}, we quickly learned that the vast
majority of the planet candidates were actually eclipsing binaries
masquerading as transiting planets \citep{latham2003}. One of the
most common imposters was an F- or G-star primary eclipsed by a late
M-dwarf secondary. Such systems produce light curves similar to
planets transiting solar-type dwarfs, because M dwarfs near the bottom
of the main sequence have roughly the same size as giant planets.
Nevertheless, they are easy to distinguish from planets, because their
masses are two orders of magnitude larger, and the reflex
orbital motion they induce in their parent stars has an amplitude of
at least several \kms.

We determined the rotational and radial velocities of HAT-TR-205-013
by cross correlation of the observed spectra against templates drawn
from a library of synthetic spectra calculated by Jon Morse for a grid
of \citet{Kurucz92} stellar atmospheres. The library grid has a
spacing of 250 K in effective temperature, $T_{\rm eff}$; 0.5 in log
surface gravity, $\log g$; and 0.5 in log metallicity relative to the
sun, [Fe/H].  For the final radial-velocity determinations we adopted
the template with the $T_{\rm eff}$, $\log g$, and $V_{\rm rot}$
values that gave the highest value for the peak of the correlation
coefficient, averaged over all the observed spectra, assuming solar
metallicity. For the correlation analysis we used  {\bf xcsao}
\citep{KurtM98} running inside the IRAF\footnote{IRAF (Image Reduction
and Analysis Facility) is distributed by the National Optical
Astronomy Observatories, which are operated by the Association of
Universities for Research in Astronomy, Inc., under contract with the
National Science Foundation.} environment.

Our very first spectrum of HAT-TR-205-013 revealed that the spectral
lines of the primary were broadened by about 30 \kms\ of rotation.  In
our experience this is a strong suggestion that the companion is a
star with enough mass to synchronize the rotation of the primary with
the orbital period. The second exposure showed that the velocity
varied by several \kms.  Additional spectra soon revealed a
spectroscopic orbit with a period that matched the photometric period
and an orbital semi-amplitude, $K$, of about 20 \kms\, implying that
the secondary was a small M dwarf. Altogether we accumulated 23
spectra yielding the radial velocities reported in Table \ref{tab1},
which were calculated using a template with $T_{\rm eff}=6250$ K,
$\log g=4.0$, [Fe/H]=0.0, and $V_{\rm rot} = 30$ \kms.  The parameters
of our orbital solution for HAT-TR-205-013 are reported in Table
\ref{tab2} and the corresponding velocity curve and observed
velocities are plotted in Figure \ref{fig3}.

Because the value of the rotational velocity is critical for our
determination of the radius of the primary star, we evaluated both the
internal precision of our determinations and possible systematic
errors resulting from uncertainties in the surface gravity and
metallicity of the primary star. Our final results for the mass and
radius of the primary star imply a surface gravity $\log g = 4.24$,
halfway between the nearest templates in our library of synthetic
spectra. Therefore we evaluated the rotational velocity at both $\log
g = 4.0$ and 4.5, and also at two metallicities, [Fe/H] = 0.0 and
$-0.5$. At each of the four combinations of $\log g$ and [Fe/H] we
ran correlations over a wide range of temperatures, 5500 to 7250 K,
and rotational velocities, 10 to 50 \kms. At each gravity and
metallicity we interpolated to find the temperature and rotation that gave
the highest value of the correlation coefficient averaged over all the
observed spectra weighted by the photon statistics. For these
experiments we did not use two of our observed spectra that had much
lower exposure levels than the other 21 observations. The results are
reported in Table \ref{tab3}. They show that the accuracy of our
rotational-velocity determination at a given $\log g$ and [Fe/H] is
not seriously limited by the scatter from the individual exposures,
despite the relatively low signal-to-noise ratio of our observed
spectra. The uncertainty in the mean rotational velocity at each
$\log g$ and [Fe/H] due to internal scatter is less than 0.3 \kms.

The results reported in Table \ref{tab3} also show that the systematic
errors due to uncertainties in the gravity and metallicity are not
serious. For example, the dependence of rotational velocity on
metallicity is rather weak, only 0.2 \kms\ for a change of 0.5 in
[Fe/H]. On the other hand, the dependence on gravity is rather
strong, about 1.0 \kms\ for a change of 0.5 in $\log g$. Fortunately,
the uncertainty in the actual gravity of the primary star is quite
small, less than 0.2 in $\log g$, so the rotational velocity
interpolated between the two gravities should be good to perhaps 0.2
\kms. Unfortunately we do not yet have a direct measure of the
metallicity of HAT-TR-205-013, but the fact that the templates with
[Fe/H] = 0.0 give a better match to the observed spectra than the
templates with [Fe/H] = $-0.5$ suggests that the metallicity is
probably within 0.5 dex of solar.  In the absence of an accurate
metallicity we adopt a value of [Fe/H] = $-0.2$, which is typical of
the solar neighborhood \citep{Nordstrom2004}. The temperature
corresponding to this metallicity that we derive from our spectra
is $T_{\rm eff} = 6295$ K, but with a systematic error that could
exceed 200 K because of the degeneracy between temperature and
metallicity in the analysis of our spectra. The alternative approach
of deriving a temperature from photometric indices is hampered by the
lack of accurate photometry and the possibility that there may be
significant reddening. The distance that we derive from the
Stefan-Boltzmann law using $R=1.27 \pm 0.04$ \rsun, $T_{\rm eff} = 6295
\pm 200$ K, $V_{\rm T} = 10.729 \pm 0.048$ (which corresponds to $V = 10.67 \pm 0.05$), and bolometric correction =
$-0.011$ \citep{flower1996} is $232 \pm 18$ pc. From our spectroscopic temperature and $\log g$, we estimate the spectral type of the primary to be F7V \citep{cox2000}.

Thus our experiments with different gravities and metallicities
suggest that the systematic errors due to uncertainties in the
template parameters may be less than 0.4 \kms. When combined in
quadrature with the internal precision estimate of 0.3 \kms, this
suggests that the total error in our final interpolated rotational
velocity could be as small as 0.5 \kms. Previous experience however, would indicate that the actual uncertainty is higher than this internal estimate. Therefore, to be conservative we adopt
an uncertainty of 1.0 \kms\ and use $V_{\rm rot} \sin i_{\rm rot} =
28.9 \pm 1.0$ \kms\ for our determination of the radius of the primary
and subsequent analysis.

One could also determine the value of $V_{\rm rot} \sin i_{\rm rot}$ by spectroscopically measuring the amplitude of the Rossiter-McLaughlin Effect during eclipse \citep{gaudi2007}. In the case of HAT-TR-205-013 the amplitude of this effect would be about 0.5 km s$^{-1}$, too small to be measured with the CfA Digital Speedometers, but potentially observable with an accuracy of a few percent by instruments such as HIRES on Keck 1.

Our approach relies on the assumption that the stellar rotation has been synchronized to the orbital period.  One way to test this assumption might be to use very precise photometry to derive a rotational period for the star. We did look for sinusoidal variations in the HAT photometry but did not find anything significant near the orbital period. We did detect a marginally significant variation with half the orbital period and amplitude of about 1 mmag, which is consistent with the the expected ellipsoidal deformation of the primary star. A second, more global test would be to use stellar models to estimate the mass and radius of the primary, but a prerequisite for such an analysis would be accurate determinations of the temperature and metallicity, presumably from high-quality spectra, and such results are not yet available.

\subsection{Follow-up KeplerCam Photometry} 

To provide a high-quality light curve for the analysis of the primary
eclipse of HAT-TR-205-013 we used KeplerCam on the 1.2-m telescope at
the Fred L.\ Whipple Observatory on Mount Hopkins, Arizona. KeplerCam
utilizes a monolithic 4K$\times$4K Fairchild 486 CCD that provides a
$23 \arcmin$ field and a pixel size of $0.34 \arcsec$. We used the
predicted eclipse times from our spectroscopic orbit to schedule
observations on the night of 22-23 October 2005. We successfully
observed a full eclipse, alternating between the Sloan $g$ and $i$ bands.

The seeing was approximately $2 \arcsec$ FWHM throughout the night,
but we deliberately defocused the telescope to get images with $3
\arcsec$ FWHM in order to keep the peak counts in the images well
below saturation. Over the course of the observations the focus of
the telescope, which depends weakly on temperature, was adjusted three
times to keep the image size near $3 \arcsec$ FWHM. Because we used
automatic guiding, the centroid of the images moved less than 3 pixels
over the duration of the observations. For readout we binned the pixels
$2\times2$, which gave a total readout time including overhead of 12
seconds. Exposure times were 30 seconds for the $g$ band and 10 seconds
for the $i$ band. The images were obtained in sequences of 3 $g$-band
exposures followed by 6 $i$-band exposures. All told, we were able to
collect a total of 297 images in the $g$ band and 588 in the $i$ band.
Some thin cirrus clouds were present at the beginning of the night,
with no noticeable degradation of the light curves. A quarter Moon
rose during the observations after the end of egress, and this
contributed to a slight increase in photometric scatter after egress.

All the images were reduced by applying an overscan correction and
then subtracting the two-dimensional residual bias pattern. After
correcting for shutter effects, we flattened each image using a
normalized set of combined twilight images. To produce the light
curve, we used the first image from each filter in an observing
sequence as our astrometric reference for identifying the same stars
in subsequent images. We then determined the relative shift between
images to relocate each star in the following images. We measured the
flux of each star in a $6.7 \arcsec$ circular aperture around the
position derived from the astrometric fit using {\bf daophot/phot} within
IRAF \citep{tody1986,tody1993}. We estimated the sky in an annulus
around each star with inner and outer radii of $9.4 \arcsec$ and $13.4
\arcsec$ using the sigma-rejection mode.

We iteratively selected comparison stars by removing any that showed
unusual noise or variability in their differential light curves,
obtaining 21 comparison stars for the $g$ band and 37 for the $i$
band. Based on a weighted mean flux from the comparison stars, we
calculated an extinction correction, and then applied this to each
comparison star. The typical rms residual for both bands was 1.5 mmag,
a value that increased slightly toward the end of the night due to the
increased sky brightness from the rising Moon. The main contributor of
noise to the photometry from individual images was atmospheric
scintillation, which accounted for over $65\%$ of our calculated rms
\citep{young1967}. The individual photometric measurements in the $g$ and 
$i$ bands are reported in Tables \ref{tab4} and \ref{tab5}, respectively.

\section{LIGHT-CURVE ANALYSIS}

Our KeplerCam light curves for HAT-TR-205-013 give good coverage of
the eclipse centered at HJD $2453666.747 \pm 0.001$, with more than two
hours of coverage both before the start of ingress and after the end of
egress. The eclipse itself lasts about 3 hours between first and
fourth contact and is about 2 percent deep, in agreement with the
discovery light curve from HAT.  Our KeplerCam light curves clearly
show the effects of limb darkening on eclipse shape as a function of
wavelength: the $i$-band light curve is slightly shallower and possesses
a flatter bottom than the $g$-band curve. The portions of the light
curves before and after eclipse showed a slight drift, which we
removed with a linear fit. The resulting light curves are plotted
in Figure \ref{fig4}.

From our fits to the KeplerCam light curves we derive transit centers
of $2453666.7465 \pm 0.0005$ and $2453666.7473 \pm 0.0005$ in the $g$ and
$i$ bands, respectively. When we combine these transit times with the
results from the HAT photometry, we get the ephemeris $2453666.74748
\pm 0.00018 + N \times 2.230736 \pm 0.000010$.

Knowing the duration and depth of the eclipse, together with the
orbital period, we were able to derive rough values for the ratio of
semimajor axis and radius of the secondary to the radius of the
primary star using the relations
\begin{equation}\label{eq:10}
\frac{a}{R_{\rm A}} \simeq \frac{1}{\pi} \left(\frac{P}{\Delta t_{\rm tr}}\right)
\end{equation}
\begin{equation}
\frac{R_{\rm B}}{R_{\rm A}} \simeq \sqrt{\delta}
\end{equation}
where $a$ is the orbital semimajor axis, $R_{\rm A}$ and $R_{\rm B}$
are the primary and secondary stars' radii, $P$ is the orbital period,
$\Delta t_{\rm tr}$ is the transit length, and $\delta$ is the
transit depth in relative flux. For our values of $P \sim 2.23$ days,
$\Delta t_{\rm tr} \sim 2.5$ hours and $\delta \sim 0.02$, we obtained
$a/R_{\rm A} \sim 5.93$ and $R_{\rm B}/R_{\rm A} \sim 0.133$ from this
first approximation.

We then built a two dimensional grid of light-curve fits in $a/R_{\rm
A}$ and $R_{\rm B}/R_{\rm A}$, centered on the rough values obtained
above. The impact parameter $b$ was introduced as a third dimension in
the grid, varying from $b=0$ (central transit) to $b=1+R_{\rm
B}/R_{\rm A}$ (grazing transit). The grid steps were $0.002$ in
$a/R_{\rm A}$, $0.00001$ in $R_{\rm B}/R_{\rm A}$, and $0.001$ in $b$. We
generated synthetic light-curves for each combination of parameters
using the routines provided by \citet{mandel2002}, together with the quadratic
limb-darkening coefficients derived by \citet{claret2004}, using the
temperature and surface gravity for the primary we had derived
spectroscopically. We adopted solar values for the metallicity and
surface turbulence of the primary star. The exact coefficients we used were $u_1=0.4238$ and $u_2=0.3250$ in the $g$, and $u_1=0.1814$ and $u_2=0.3723$ in the $i$ band. Uncertainties in our values of the limb-darkening coefficients, largely a result of the uncertainty in the spectroscopic temperature measurement, had a negligible effect on our final results. 

To identify the synthetic light curve that gave the best fit to the
actual data, we looked for the fit with the smallest value of
$\chi^2$, inspecting the $\chi^2$ contours to ensure that we had found
the global minimum. We also inspected plots of the fits to the data
as a qualitative check. We applied this procedure to both the $g$ and
$i$ light curves. The results are summarized in Table \ref{tab6}.

To calculate the uncertainties for our values of $a/R_{\rm A}$, $R_{\rm B}/R_{\rm A}$, and $b$, we performed additional fits wherein we fixed one of the parameters at a value differing slightly from the best-fit value, and fit for the other two parameters. We changed this fixed parameter until the eventual fit achieved a $\chi^2$ that was 1-$\sigma$ away from our best fit parameters' $\chi^2$. We then used the difference between this 1-$\sigma$ value and our best fit value as the uncertainty for that particular parameter. We examined the correlation between the three different parameters by making a final fit to the data using the Levenberg-Marquardt method -- a combination of the Inverse-Hessian and Steepest Descent methods \citep{press1992}. This allowed us to compute the covariance matrix for the three fit parameters, and thus the correlation coefficients for the parameters. For all three, the correlation with any of the other two was always smaller than 0.5 in both bands. While this is not negligible, we found that the contribution to the total estimated error was minor. Of course, the Levenberg-Marquardt method can also be used to calculate the uncertainties in $a/R_{\rm A}$, $R_{\rm B}/R_{\rm A}$, and $b$. However, the uncertainties that we derived from our grid analysis were substantially larger. To be conservative, we adopted the grid uncertainties instead of the Levenberg-Marquardt uncertainties. Figure \ref{fig5} shows contour plots of $\chi^2$ for our fits. Table \ref{tab6} reports the final fitting results and errors, and Table \ref{tab7} lists the correlation coefficients. The KeplerCam data and best-fit light curves are shown in Figure \ref{fig4}. Simultaneous fitting of the light curves in both bands produced fits with inferior $\chi^2$ statistics, a result of the slight difference in $R_{\rm B}/R_{\rm A}$ in the two bands. We therefore use a simple average of the light curve results as our adopted values for the parameters in Table \ref{tab6}.

\subsection{MASSES AND RADII FOR HAT-TR-205-013}

Using the values that we have measured from the eclipse light curves,
as well as the observed spectroscopic orbit parameters, we may
restrict the location of HAT-TR-205-013 A and B on the mass-radius diagram to unique curves described
completely by these observables. To emphasize this, we have written
the observable values in brackets in the following derivation.

To begin, we use Newton's revised version of Kepler's Third Law and
the spectroscopic mass function from the orbital solution:
\begin{equation}
[P]^2 = \frac{4 \pi^{2}}{G\left(M_{\rm A} + M_{\rm B}\right)} \ a^{3}
\end{equation}
\begin{equation}
M_{\rm B} = \left(\frac{[P]}{2 \pi G}\right)^{1/3} \ \frac{[K_{\rm A}]}{\sin i_{\rm orb}} 
\left(M_{\rm A} + M_{\rm B}\right)^{2/3} \sqrt{1-[e]^2}
\end{equation}
where $[K_{\rm A}]$ is the semi-amplitude of the spectroscopic
orbit for the primary in km s$^{-1}$, $[e]$ the eccentricity of the
orbit, and $G$ is the Gravitational Constant. With these two equations,
the two unknowns ($M_{\rm A}$ and $M_{\rm B}$) may be found:
\begin{eqnarray}
M_{\rm A} & = & \frac{4 \pi^2}{G} \ \frac{a^3}{[P]^2} \ \left(1 - \frac{[P] \ [K_{\rm A}] \ \sqrt{1-[e]^2} }{ 2 \pi a \sin i_{\rm orb}} \right) \\
M_{\rm B} & = & \frac{2 \pi}{G} \ \frac{a^2}{[P]} \ \frac{[K_{\rm A}]}{\sin i_{\rm orb}} \sqrt{1-[e]^2}
\end{eqnarray}

Both the orbital inclination and the semimajor axis can be expressed in
terms of the observables $[a/R_{\rm A}]$, $[R_{\rm B}/R_{\rm
A}]$ and $[b]$, and the unknown stellar radii $R_{\rm A}$ and $R_{\rm B}$:

By taking our value for the impact parameter $[b]$, which measures the projected separation of the primary and secondary stars at the mid-point of the eclipse, the orbital inclination can be expressed in terms of the observables $[a/R_{\rm A}]$ and $[b$]:
\begin{equation}
[b] = [a/R_{\rm A}] \cos i_{\rm orb}
\end{equation}
\begin{equation}
\sin i_{\rm orb} = \left(1 - \frac{[b]^2}{[a/R_{\rm A}]^2}\right)^{1/2}
\end{equation}

Similarly, the semimajor axis may be written in terms of the observables $[a/R_{\rm A}]$, $[R_{\rm B}/R_{\rm A}]$ and $[b]$, and the unknown stellar radii $R_{\rm A}$ and $R_{\rm B}$:
\begin{equation}
a = [a/R_{\rm A}] \ R_{\rm A} = \left(\frac{[a/R_{\rm A}]}{[R_{\rm B}/R_{\rm A}]} \right) \ R_{\rm B}
\end{equation}

Substituting these values into Eq.(5) and (6) gives expressions for the mass of each component as functions of only the observables and the respective stellar radii:
\begin{eqnarray}
M_{\rm A} & = & \frac{4 \pi^{2}}{G [P]^2} \ [a/R_{\rm A}]^3 \  \left(1 - \frac{[P] \ [K_{\rm A}] \ \sqrt{1-[e]^2}}{2 \pi {\left(1-[b]^2/[a/R_{\rm A}]^2\right)}^{1/2}\ [a/R_{\rm A}]\ R_{\rm A}}\right) \ R_{\rm A}^3 \\
M_{\rm B} & = & \frac{2 \pi}{G [P]} \left(\frac{[a/R_{\rm A}]}{[R_{\rm B}/R_{\rm A}]}\right)^2 \frac{[K_{\rm A}]\ \sqrt{1-[e]^2}}{\left(1-[b]^2/[a/R_{\rm A}]^2\right)^{1/2}}\ R_{\rm B}^2
\end{eqnarray}

Therefore, solely from what we are able to measure using the eclipse light curves and spectroscopy, we may confine HAT-TR-205-013 B to a single curve on the mass-radius diagram that goes as $M_{\rm B} /R_{\rm B}^2 = \rm{constant}$. This constant is directly proportional to the surface gravity of the object, which has been noted previously by \cite{southworth2004}. In our case, the quality of our photometry allows us to measure this constant extremely accurately; the uncertainty region around this surface gravity line is on the order of the line width itself when it is plotted.

Through the assumed synchronization of the primary's rotation to the orbital period we may locate HAT-TR-205-013 B on this gravity determination curve, but it is important to note that the line itself is defined without having made any assumptions about the system. Indeed, it is possible to calculate a similar gravity curve for any eclipsing system, or for a system containing a transiting planet. All that is required is a good-quality light curve and spectroscopic orbit. This has previously been done for the case of a transiting planet by \cite{winn2006}.

To place HAT-TR-205-013 on the mass-radius diagram more specifically, we note that that the measured eccentricity of our spectroscopic orbit for HAT-TR-205-013 is indistinguishable from circular, and therefore, for the reasons described in the Introduction, we assume that the spin axes of both stars have been aligned with the orbital normal and that the rotation of both stars has been synchronized to the orbital period. This allows us to use the observed rotational line broadening of the primary to solve for the radius of the primary in linear units, which in turn allows us to convert the orbital size and secondary radius into linear units from the values of $[a/R_{\rm A}]$ and $[R_{\rm B}/R_{\rm A}]$ derived from the light curves.

Using the assumption of synchronization, and that $i_{\rm orb} = i_{\rm rot}$, we see by inspection that 
\begin{eqnarray}
R_{\rm A} &=& \frac{[P]}{2\pi}\ \frac{[V_{\rm rot} \sin i_{\rm rot}]}
{\sin i_{\rm orb}}\\
R_{\rm B} &=& \frac{[P]}{2\pi}\ \frac{[V_{\rm rot} \sin i_{\rm rot}]}
{\sin i_{\rm orb}} [R_{\rm B}/R_{\rm A}]
\end{eqnarray} 
where $[V_{\rm rot} \sin i_{\rm rot}]$ is the projected rotational broadening of 
the primary derived from its observed spectra. We may now substitute in Eq.(8) for $\sin i_{\rm orb}$ to get both radii in terms of our observables:
\begin{eqnarray}
R_{\rm A} & = & \frac{[P]}{2\pi \left(1-[b]^2/[a/R_{\rm A}]^2\right)^{1/2}}\ [V_{\rm rot} \sin i_{\rm rot}]\\
R_{\rm B} & = & \frac{[P]}{2\pi \left(1-[b]^2/[a/R_{\rm A}]^2\right)^{1/2}}\ [R_{\rm B}/R_{\rm A}]\ [V_{\rm rot} \sin i_{\rm rot}]
\end{eqnarray}  

By combining these two statements with Eq.(9) and (10), we arrive at expressions for the masses of each component in terms of just the observable quantities:
\begin{eqnarray}
M_{\rm A} & = & \frac{[P]}{2\pi G} \frac{[a/R_{\rm A}]^3}{\left(1-[b]^2/[a/R_{\rm A}]^2\right)^{3/2}}\ \left(1 - \frac{[K_{\rm A}]\sqrt{1-[e]^2}}{[a/R_{\rm A}] [V_{\rm rot} \sin i_{\rm rot}]}\right)\ [V_{\rm rot} \sin i_{\rm rot}]^3\\
M_{\rm B} & = & \frac{[P]}{2 \pi G} \frac{[a/R_{\rm A}]^2}{\left(1-[b]^2/[a/R_{\rm A}]^2\right)^{3/2}}\ [K_{\rm A}]\ \sqrt{1-[e]^2}\ [V_{\rm rot} \sin i_{\rm rot}]^2
\end{eqnarray}  

The results for the masses and radii for both components of
HAT-TR-205-013 are presented in Table \ref{tab8}. The errors were
estimated using Monte-Carlo simulations and were compared with the
results of formal error propagation, including the correlation
coefficients derived from the light-curve fits. Both approaches
delivered similar results. The mass and radius obtained for the
primary star are essentially the same for both the $g$ and $i$ light
curves, but the mass and radius for the secondary differ by 0.8 and 3
percent, respectively. This radius difference between the two light
curves is close to 1-$\sigma$, and may be due to uncertainties in the
limb-darkening coefficients. Our adopted values are based on the average values of the light curve parameters.

\section{DISCUSSION}

In Figure \ref{fig6} we plot our mass and radius for the M-dwarf secondary HAT-TR-205-013 B on a mass-radius diagram, together with isochrones for ages of 0.5 and 5 Gyr from \cite{baraffe1998}. We also plot the results for 11 M dwarf secondaries from the sample of OGLE planetary candidates analyzed by \citet{bouchy2005,pont2005a,pont2005b,pont2006} and listed in Table \ref{tab9}. For the systems OGLE-TR-34 \citep{bouchy2005}, OGLE-TR-120 \citep{pont2005b}, and the low mass systems OGLE-TR-122 \citep{pont2005a} and OGLE-TR-123 \citep{pont2006} the authors had to use stellar models to to estimate the masses and radii of the primaries without the assumption of synchronization, as synchronization implied masses and radii that were inconsistent with the spectroscopic observations. For the other seven systems, they were able to assume synchronization and to derive the radius of the primary from the observed rotational line broadening. In general the agreement between the OGLE results and the \citet{baraffe1998} isochrones looks promising, but the observational uncertainties are still too large to allow a critical test of the theoretical models. The OGLE systems are all much fainter than HAT-TR-205-013, which presents significant challenges for both the spectroscopic and photometric follow-up observations. Spectroscopy with the resolution and signal-to-noise ratio suitable for determining accurate values for rotational broadening requires time on large telescopes, and photometry for high-quality light curves also requires large telescopes to achieve the needed photon statistics. Eclipsing binaries identified by wide-angle surveys are much brighter and therefore less challenging on both counts.

Our value for the radius of the M-dwarf secondary in HAT-TR-205-013 lies 11 percent, about 3-$\sigma$, above the theoretical isochrones. This divergence is further reinforced by Eq.(11), which, as has been previously noted, restricts the position of HAT-TR-205-013 B to lie on a single line that is determined by the surface gravity of the object. This gravity curve does not rely on any prior assumptions about the HAT-TR-205-013 system, nor does it depend upon our measured value of $V_{\rm rot} \sin i_{\rm rot}$, which is the biggest contributor of uncertainty to our final results. We use the assumption of synchronization and the spectroscopically measured $V_{\rm rot} \sin i_{\rm rot}$ to place HAT-TR-205-013 B at a specific location along the curve, but it is important to note that in the region that we find HAT-TR-205-013 B, the curve of allowable locations runs nearly parallel to the theoretical models. This is illustrated in Figure \ref{fig6} by the red line that passes through our point for HAT-TR-205-013 B. Thus the conclusion that the theoretical models predict a radius for HAT-TR-205-013 B that is too small by about 10 percent is on much firmer ground than the error bar in the observed radius might suggest. Indeed, it would require a 6-$\sigma$ difference in $V_{\rm rot} \sin i_{\rm rot}$ to place HAT-TR-205-013 B onto the Baraffe models. 

\begin{figure}
\plotone{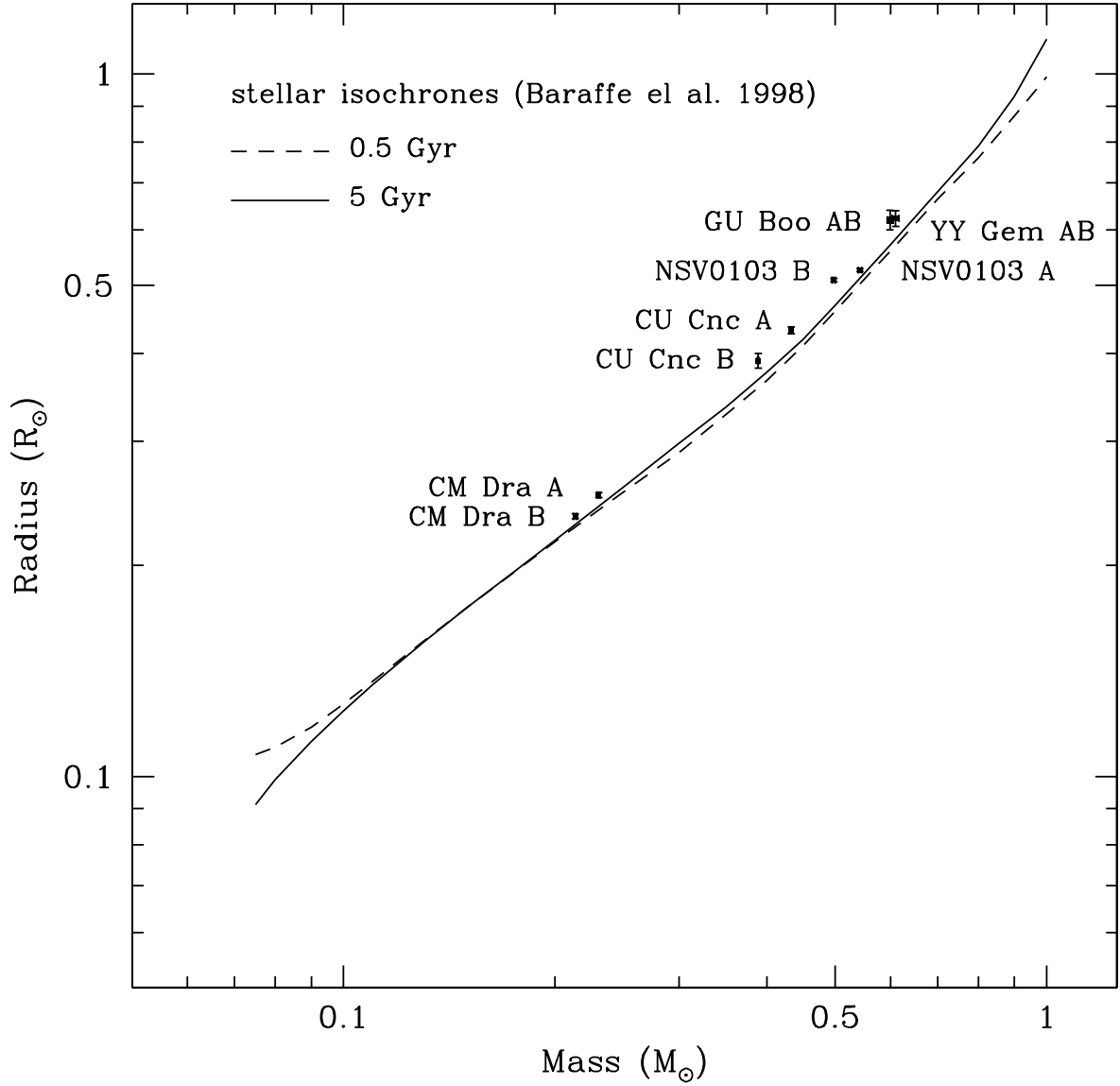}
\caption{The mass-radius diagram for 10 stars in 5 double-lined eclipsing
binaries each composed of two M dwarfs, and with errors better than 3
percent. \label{fig1}}
\end{figure}

\begin{figure}
\plotone{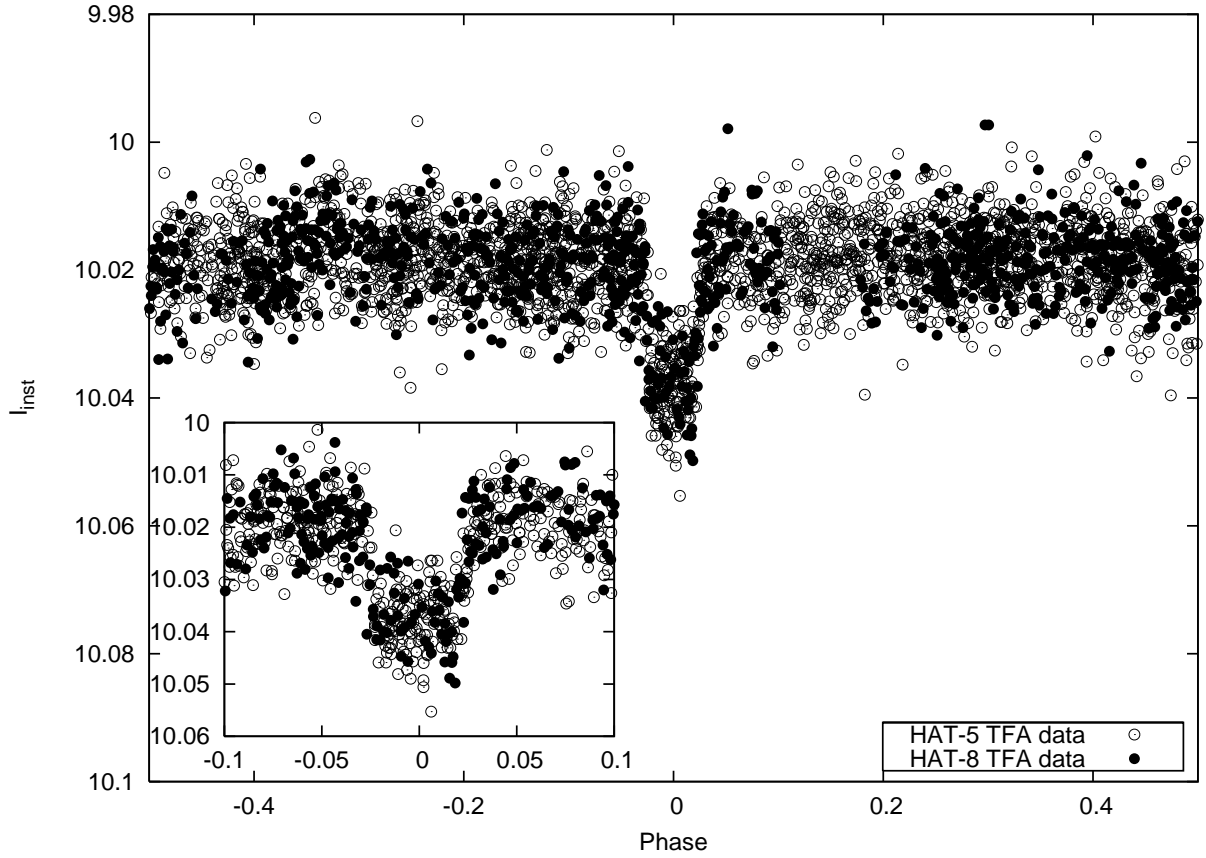}
\caption{The phase-folded HATnet light curve for HAT-TR-205-013. \label{fig2}}
\end{figure}

\begin{figure}
\plotone{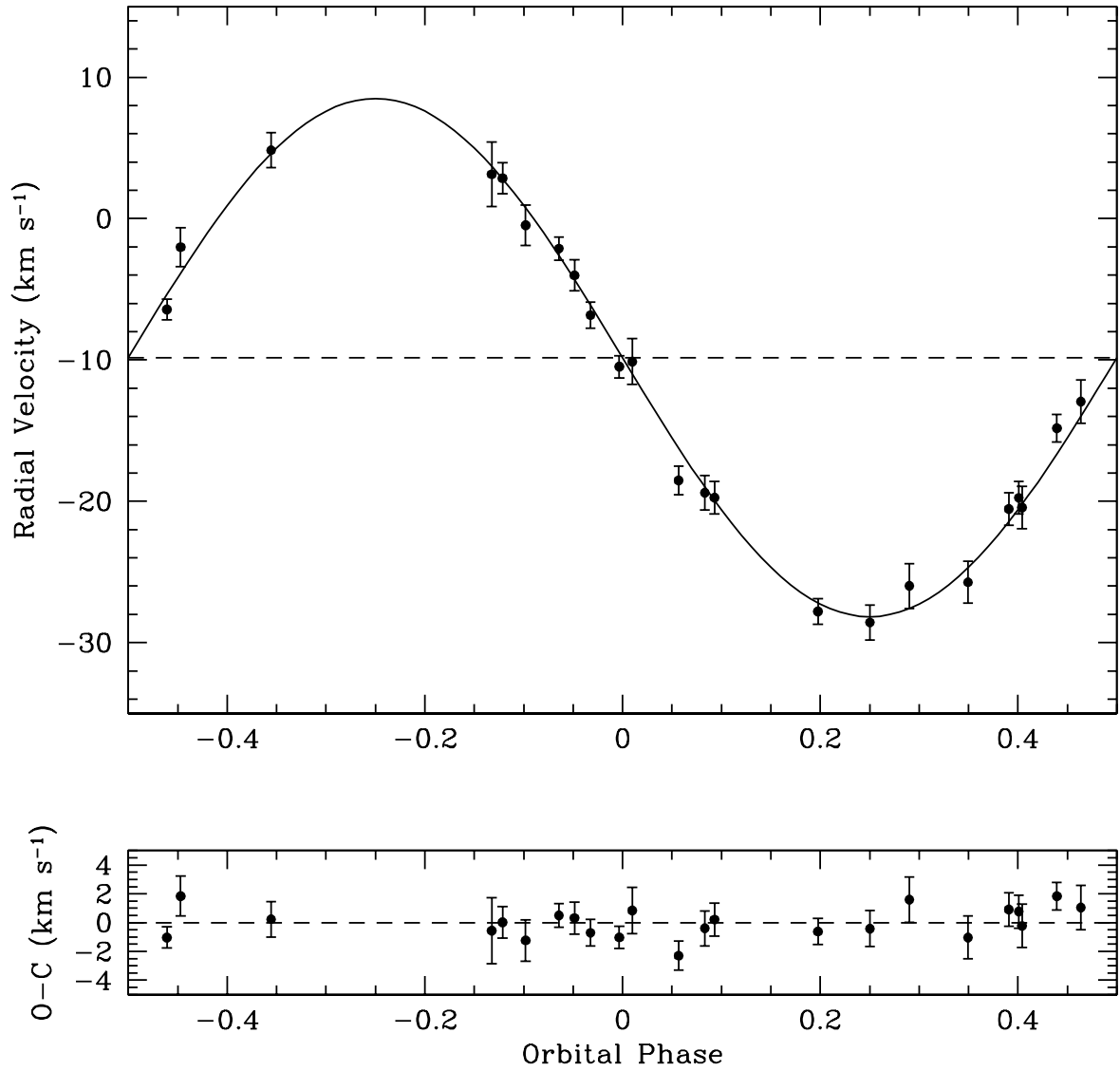}
\caption{The velocity curve for our orbital solution for HAT-TR-205-013, 
together with the individual observed velocities.  The lower panel
shows the O-C velocity residuals from the orbital solution. \label{fig3}}
\end{figure}

\begin{figure}
\plotone{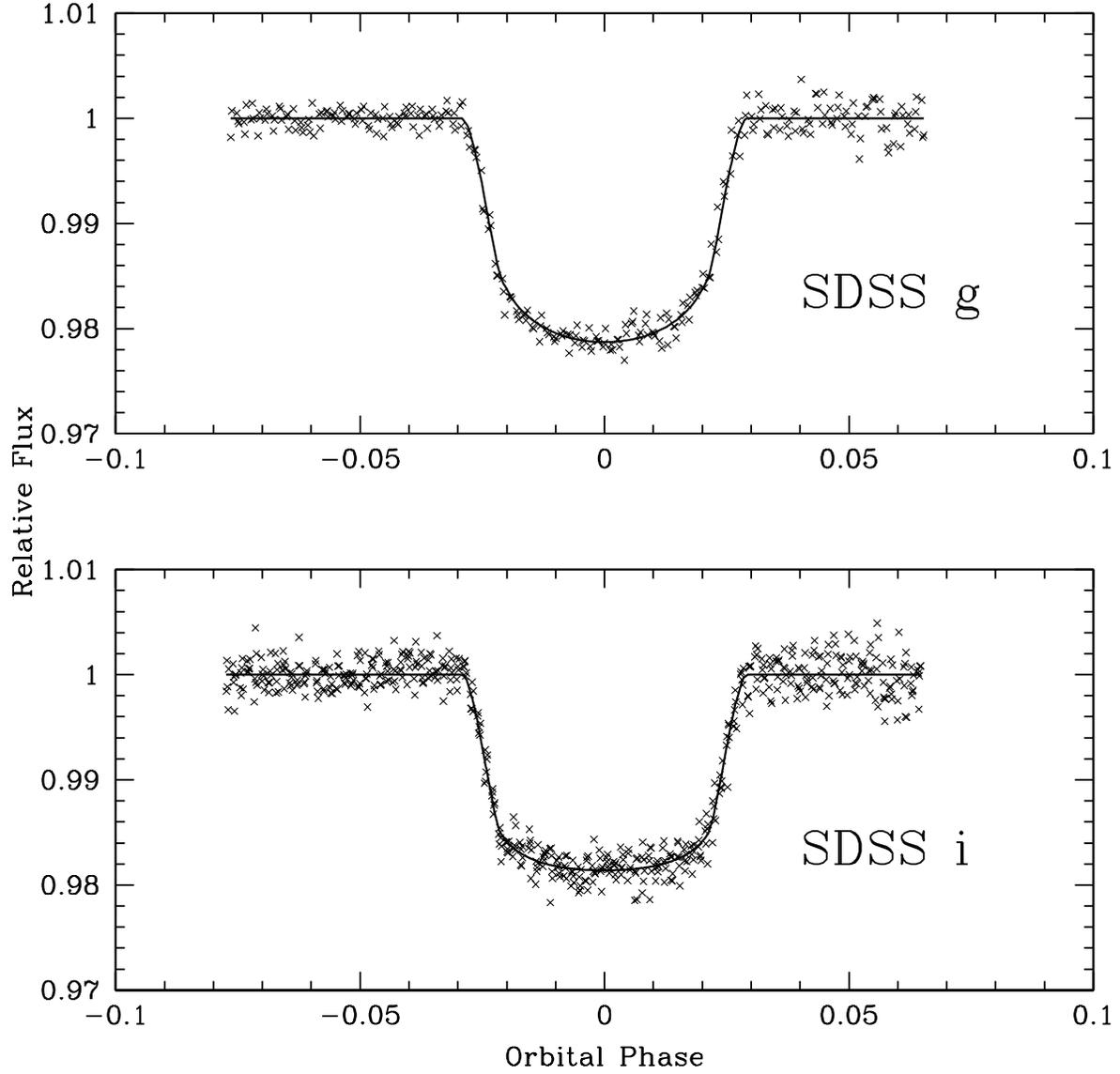}
\caption{KeplerCam light curves for HAT-TR-205-013 in the SDSS $g$ and $i$ bands. Continuous lines show the best fit synthetic light curves for each.
\label{fig4}}
\end{figure}

\begin{figure}
\plotone{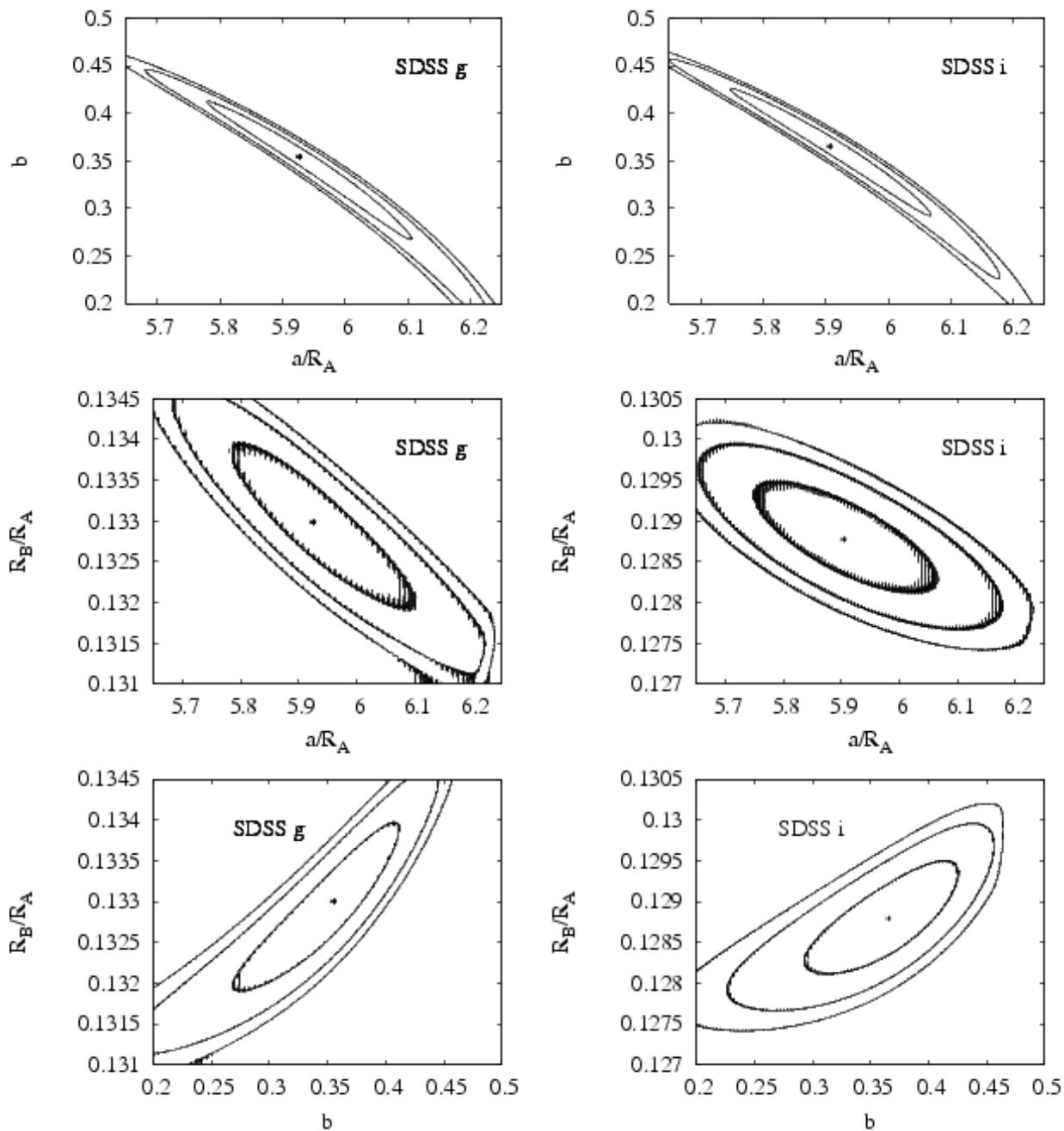}
\caption{Contours of $\chi^2$ for the results from fits to the light
curves in the $g$-band (left panels) and $i$-band (right panels).
For each band the three panels show the projections onto the three
possible planes involving $b$, $a/R_{\rm A}$, and $R_{\rm B}/R_{\rm A}$.
The 1-$\sigma$, 2-$\sigma,$ and 3-$\sigma$ contours are
plotted. \label{fig5}}
\end{figure}

\begin{figure}
\plotone{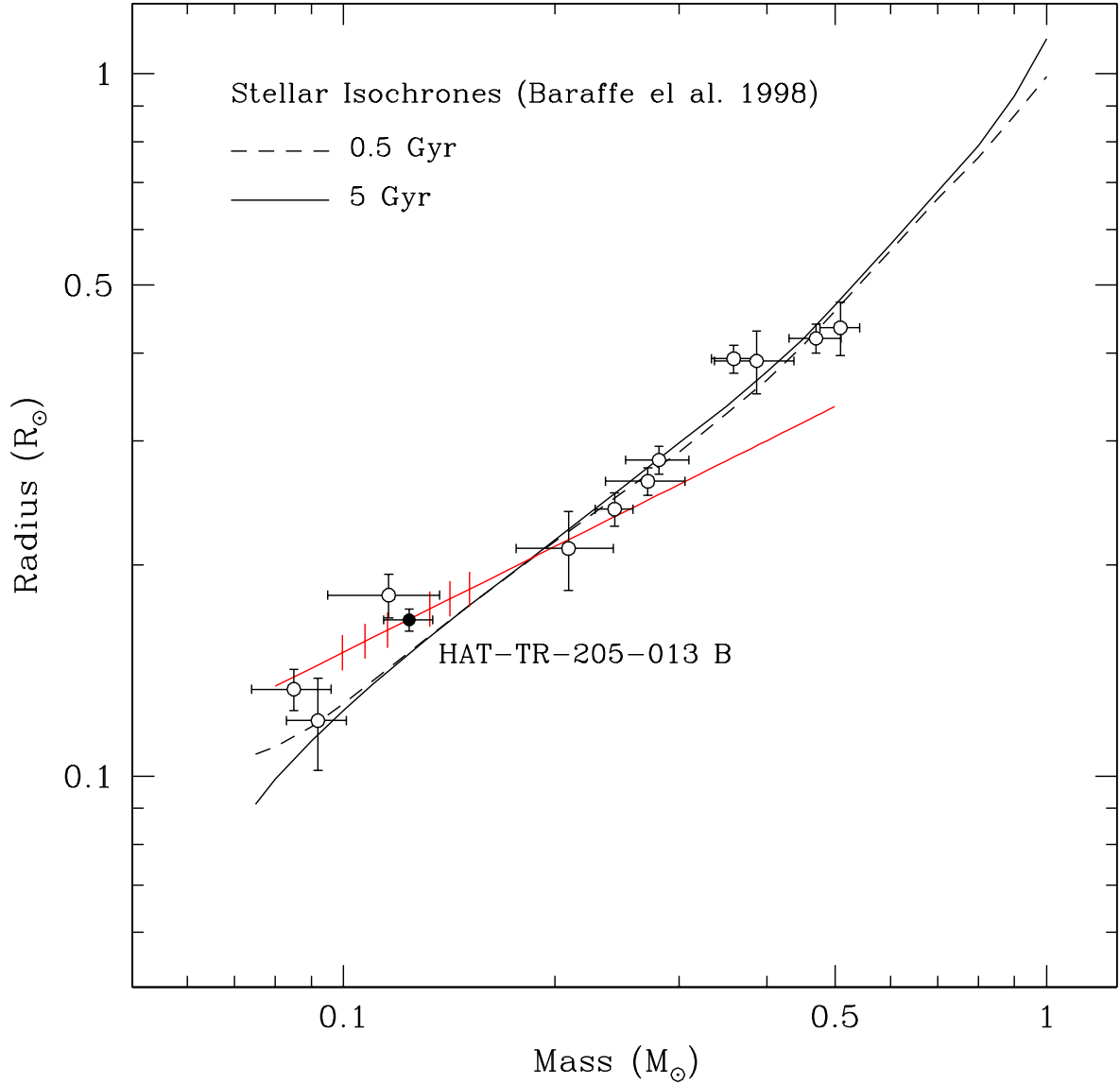}
\caption{The mass-radius diagram for M dwarfs in single-lined
eclipsing binaries. The M dwarfs from \cite{pont2005a,pont2005b,pont2006} are
plotted as open circles. The red line passing through the point for
HAT-TR-205-013 B shows the constraint imposed on its location by Eq.(11) and our observed quantities, without making any explicit assumptions (such as synchronization) about the system. Assuming synchronization, the hash marks on the line show the effect that differences of $\pm$ 1, 2, \& 3 \kms \ in $V_{\rm rot}$ have on our final results. \label{fig6}}
\end{figure}

Our result for HAT-TR-205-013 B supports the suggestion from the results for double-lined eclipsing binaries plotted in Figure \ref{fig1} that the models predict radii for M dwarfs that are too small by up to 10 percent. This discrepancy has been noted before, for example by \citet{torres2002} in the case of YY Gem. \citet{torres2006} raised the issue of whether short-period eclipsing binaries are representative of isolated field stars and wide binaries where tidal forces are negligible. They suggested that the rapid rotation of the stars in these systems caused by tidal synchronization might give rise to enhanced magnetic activity, thus decreasing the efficiency of energy transport in the convective envelopes and leading to inflated stellar radii. For low mass stars, this effect is examined in more detail by \cite{lopez2007}.

In the case of HAT-TR-205-013, we see no evidence in the photometry of star spots on the primary star, which would be tell-tale indicators of enhanced stellar magnetic activity. Though HAT-TR-205-013 A is rapidly rotating, the lack of magnetic ativity is not suprising, given its spectral type (F7). The star's outer convective layer is relatively shallow, and it is not unusual for rapidly rotating stars of this type to lack strong magnetic activity \citep{torres2006}.

In some instances it may be possible to independently determine the rotational period of the primary through high-quality light curves used to definitively identify photometric variation outside of eclipse. This would serve as a check to the assumption of tidal synchronization in the system. 

In future papers we will present the results for several additional single-lined eclipsing binaries with circularized orbits.

\acknowledgments
We thank Joe Zajac, Perry Berlind, and Mike Calkins
for obtaining some of the spectroscopic observations; Bob Davis for
maintaining the database for the CfA Digital Speedometers; and John
Geary, Andy Szentgyorgyi, Emilio Falco, Ted Groner, and Wayne Peters
for their contribution to making KeplerCam such an effective
instrument for obtaining high-quality light curves. TGB thanks the
Harvard University Origins of Life Initiative for support. GK thanks the support of OTKA K-60750. 
The HATnet project is supported by NASA Grant NNG04GN74G. This research was supported in part by the Kepler Mission under NASA Cooperative Agreement NCC2-1390.

\clearpage
\begin{deluxetable}{rrc}
\tablewidth{0pc}
\tablenum{1}
\tablecaption{\sc Individual Radial Velocities \label{tab1}}
\tablehead{
\colhead{HJD}                    &
\colhead{$V_{\rm rad}$}          &
\colhead{$\sigma (V_{\rm rad)}$} \\
\colhead{(days)}                 &
\colhead{(\kms)}                 &
\colhead{(\kms)}
}
\startdata
2453034.45642 &  $-$2.02 &  1.38 \\
2453035.47574 & $-$10.11 &  1.61 \\
2453035.58018 & $-$18.52 &  1.01 \\
2453036.48778 & $-$12.93 &  1.53 \\
2453037.46565 &  $-$0.47 &  1.43 \\
2453037.61215 &  $-$6.83 &  0.91 \\
2453038.46413 & $-$25.72 &  1.48 \\
2453038.57874 & $-$19.76 &  1.15 \\
2453040.47360 & $-$28.58 &  1.24 \\
2453042.58686 & $-$27.79 &  0.91 \\
2453043.58338 &    +4.84 &  1.23 \\
2453044.58422 & $-$19.73 &  1.14 \\
2453045.57911 &  $-$6.42 &  0.73 \\
2453046.46373 &  $-$2.11 &  0.80 \\
2453046.60000 & $-$10.47 &  0.77 \\
2453047.50881 & $-$20.45 &  1.49 \\
2453047.58731 & $-$14.83 &  0.95 \\
2453543.94910 &  $-$4.01 &  1.10 \\
2453658.69572 & $-$20.53 &  1.16 \\
2453659.75967 &    +3.14 &  2.28 \\
2453659.78398 &    +2.85 &  1.09 \\
2453660.70213 & $-$25.98 &  1.57 \\
2453664.70202 & $-$19.39 &  1.21 \\
\enddata
\end{deluxetable}

\clearpage
\begin{deluxetable}{lc}
\tablewidth{0pt}
\tablenum{2}
\tablecaption{\sc Spectroscopic Orbital Parameters \label{tab2}}
\tablehead{
\colhead{Parameter} &
\colhead{Value}}
\startdata
$P$ (days)               & $ 2.23072  \pm 0.00005 $ \\
$\gamma$ (\kms)          & $ -9.83    \pm 0.30    $ \\
$K$ (\kms)               & $ 18.33    \pm 0.47    $ \\
$e$                      & $ 0.012    \pm 0.021   $ \\
$\omega$ ($\arcdeg$)     & $ 143      \pm 90      $ \\
Epoch (HJD)              & $ 2,453,198.61 \pm 0.56    $ \\
Nobs                     & $ 23                   $ \\
$O-C$ rms (\kms)         & $ 1.06                 $ \\
$f(M)$ ($\msun^3$)       & $ 0.00142  \pm 0.00023 $ \\
$a_{\rm A} \sin i$ (Gm)  & $ 0.562    \pm 0.030   $ \\
\enddata
\end{deluxetable}

\clearpage
\begin{deluxetable}{ccccc}
\tablewidth{0pt}
\tablenum{3}
\tablecaption{\sc Rotational Velocity Results  \label{tab3}}
\tablehead                       {
\colhead{$\log g$,[Fe/H]}        &
\colhead{$T_{\rm eff}$}          &
\colhead{$<V_{\rm rot}>$}        &
\colhead{$\sigma(<V_{\rm rot}>)$} &
\colhead{Correlation}               \\
\colhead{}                       &
\colhead{(K)}                    &
\colhead{(\kms)}                 &
\colhead{(\kms)}                 &
\colhead{Coefficient}                       }
\startdata
4.0,0.0    & 6340 & 29.4 & 0.25 & 0.826 \\
4.5,0.0    & 6540 & 28.4 & 0.24 & 0.823 \\
4.0,$-0.5$ & 5960 & 29.2 & 0.21 & 0.821 \\
4.5,$-0.5$ & 6150 & 28.2 & 0.24 & 0.816 \\
Adopted:   &      &      &      &       \\
4.24,$-0.2$& 6295 & 28.9 & 1.0  &       \\
\enddata
\end{deluxetable}

\clearpage
\begin{deluxetable}{cc}
\tablewidth{0pt}
\tablenum{4}
\tablecaption{\sc $g$ Band Photometry \label{tab4}}
\tablehead{
\colhead{HJD}   &
\colhead{Flux}}
\startdata
 2453666.575985 & 1.00054 \\
 2453666.576472 & 0.99856 \\
 2453666.576946 & 1.00108 \\
 2453666.579226 & 1.00084 \\
 2453666.579712 & 1.00008 \\
 2453666.580198 & 0.99985 \\
 2453666.582501 & 0.99998 \\
 2453666.582976 & 0.99882 \\
 2453666.583474 & 1.00159 \\
\enddata
\tablecomments{
Table 4 is presented in its entirety in the electronic edition of the
Astrophysical Journal. A portion is shown here for guidance regarding
its form and content.\\
column (1): Heliocentric Julian Date,\\
column (2): Normalized instrumental flux.}
\end{deluxetable}

\clearpage
\begin{deluxetable}{cc}
\tablewidth{0pt}
\tablenum{5}
\tablecaption{\sc $i$ Band Photometry \label{tab5}}
\tablehead{
\colhead{HJD}   &
\colhead{Flux}}
\startdata
 2453666.574226 & 0.99951 \\
 2453666.574469 & 0.99803 \\
 2453666.574724 & 1.00096 \\
 2453666.574967 & 0.99846 \\
 2453666.575233 & 0.99628 \\
 2453666.575488 & 1.00014 \\
 2453666.577432 & 1.00062 \\
 2453666.577698 & 0.99886 \\
 2453666.577965 & 0.99941 \\
 2453666.578208 & 0.99613 \\
 2453666.578474 & 0.99957 \\
 2453666.578728 & 0.99962 \\
 2453666.580719 & 0.99837 \\
 2453666.580974 & 0.99976 \\
 2453666.581228 & 0.99761 \\
 2453666.581494 & 1.00025 \\
 2453666.581772 & 1.00113 \\
 2453666.582027 & 0.99823 \\
\enddata
\tablecomments{
Table 5 is presented in its entirety in the electronic edition of the
Astrophysical Journal. A portion is shown here for guidance regarding
its form and content.\\
column (1): Heliocentric Julian Date,\\
column (2): Normalized instrumental flux.}
\end{deluxetable}

\clearpage
\begin{deluxetable}{cccc}
\tablewidth{0pt}
\tablenum{6}
\tablecaption{\sc Light-Curve Fit Results \label{tab6}}
\tablehead{
\multicolumn{1}{c}{Parameter}&
\multicolumn{1}{c}{$g$ Band}&
\multicolumn{1}{c}{$i$ Band}&
\multicolumn{1}{c}{Adopted}
}
\startdata
$a/R_{\rm A}$         & $5.93   \pm 0.15$   & $5.91   \pm 0.16$   & $5.92   \pm 0.11$\\
$R_{\rm B}/R_{\rm A}$ & $0.1330 \pm 0.0010$ & $0.1288 \pm 0.0007$ & $0.1309 \pm 0.0006$\\
$b$                   & $0.36   \pm 0.06$   & $0.37   \pm 0.07$   & $0.365  \pm 0.046$\\
\enddata
\end {deluxetable}

\clearpage
\begin{deluxetable}{ccc}
\tablewidth{0pt}
\tablenum{7}
\tablecaption{\sc Correlation Coefficients \label{tab7}}
\tablehead{
\multicolumn{1}{c}{Coefficient}&
\multicolumn{1}{c}{$g$ Band}&
\multicolumn{1}{c}{$i$ Band}}
\startdata
($a/R_{\rm A}$,$R_{\rm B}/R_{\rm A}$)  & $0.28$  & $0.27$  \\
($a/R_{\rm A}$,$b$)                    & $-0.21$ & $-0.42$ \\
($R_{\rm B}/R_{\rm A}$,$b$)            & $0.04$  & $0.01$  \\
\enddata
\end {deluxetable}

\clearpage
\begin{deluxetable}{cccc}
\tablewidth{0pt}
\tablenum{8}
\tablecaption{\sc Physical Parameters for HAT-TR-205-013 \label{tab8}}
\tablehead{
\multicolumn{1}{c}{Parameter}&
\multicolumn{1}{c}{$g$ Band}&
\multicolumn{1}{c}{$i$ Band}&
\multicolumn{1}{c}{Adopted}}
\startdata
$M_{\rm A}$ (\msun) & $1.04 \pm 0.14$ & $1.03 \pm 0.14$ & $1.04 \pm 0.13$ \\
$R_{\rm A}$ (\rsun) & $1.28 \pm 0.04$ & $1.28 \pm 0.04$ & $1.28 \pm 0.04$ \\
\\
$M_{\rm B}$ (\msun) & $0.124 \pm 0.011$ & $0.123 \pm 0.011$ & $0.124 \pm 0.010$ \\
$R_{\rm B}$ (\rsun) & $0.169 \pm 0.006$ & $0.164 \pm 0.006$ & $0.167 \pm 0.006$ \\
\\
$a$ (AU)            & $0.0351 \pm 0.0015$ & $0.0351 \pm 0.0015$ & $0.0351 \pm 0.0014$
\enddata
\end {deluxetable}

\clearpage
\begin{deluxetable}{lcccc}
\tablewidth{0pt}
\tablenum{9}
\tablecaption{\sc Masses and Radii for Low-Mass Stars \label{tab9}}
\tablehead{
\multicolumn{1}{l}{Name}                &
\multicolumn{1}{c}{$M$ (\msun)}         &
\multicolumn{1}{c}{$R$ (\rsun)}         &
\multicolumn{1}{c}{Type}                &
\multicolumn{1}{c}{Ref.}}
\startdata
OGLE-TR-123 B     &$ 0.085 \pm 0.011   $&$ 0.133 \pm 0.009   $& SB1 EB & 1 \\
OGLE-TR-122 B     &$ 0.092 \pm 0.009   $&$ 0.120 \pm 0.018   $& SB1 EB & 2,3 \\
OGLE-TR-106 B     &$ 0.116 \pm 0.021   $&$ 0.181 \pm 0.013   $& SB1 EB & 3 \\
HAT-TR-205-013 B  &$ 0.123 \pm 0.011   $&$ 0.167 \pm 0.007   $& SB1 EB & 13 \\
OGLE-TR-125 B     &$ 0.209 \pm 0.033   $&$ 0.211 \pm 0.027   $& SB1 EB & 3 \\
CM Dra B          &$ 0.2136 \pm 0.0010 $&$ 0.2347 \pm 0.0019 $& SB2 EB & 4,5 \\
CM Dra A          &$ 0.2307 \pm 0.0010 $&$ 0.2516 \pm 0.0020 $& SB2 EB & 4,5 \\
OGLE-TR-78 B      &$ 0.243 \pm 0.015   $&$ 0.240 \pm 0.013   $& SB1 EB & 3 \\
OGLE-TR-5 B       &$ 0.271 \pm 0.035   $&$ 0.263 \pm 0.012   $& SB1 EB & 6 \\
OGLE-TR-7 B       &$ 0.281 \pm 0.029   $&$ 0.282 \pm 0.013   $& SB1 EB & 6 \\
OGLE-TR-6 B       &$ 0.359 \pm 0.025   $&$ 0.393 \pm 0.018   $& SB1 EB & 6 \\
OGLE-TR-18 B      &$ 0.387 \pm 0.049   $&$ 0.390 \pm 0.040   $& SB1 EB & 6 \\
CU Cnc B          &$ 0.3890 \pm 0.0014 $&$ 0.3908 \pm 0.0094 $& SB2 EB & 7 \\
OGLE-BW3-V38 B    &$ 0.41 \pm 0.09     $&$ 0.44 \pm 0.06     $& SB2 EB & 8 \\
CU Cnc A          &$ 0.4333 \pm 0.0017 $&$ 0.4317 \pm 0.0052 $& SB2 EB & 7 \\
OGLE-BW3-V38 A    &$ 0.44 \pm 0.07     $&$ 0.51 \pm 0.04     $& SB2 EB & 8 \\
OGLE-TR-120 B     &$ 0.47 \pm 0.04     $&$ 0.42 \pm 0.02     $& SB1 EB & 3 \\
TrES-Her0-07621 B &$ 0.489 \pm 0.003   $&$ 0.452 \pm 0.050   $& SB2 EB & 9 \\
TrES-Her0-07621 A &$ 0.493 \pm 0.003   $&$ 0.453 \pm 0.060   $& SB2 EB & 9 \\
NSVS01031772 B    &$ 0.4982 \pm 0.0025 $&$ 0.5088 \pm 0.0030 $& SB2 EB & 10 \\
OGLE-TR-34 B      &$ 0.509 \pm 0.038   $&$ 0.435 \pm 0.033   $& SB1 EB & 6 \\
NSVS01031772 A    &$ 0.5428 \pm 0.0027 $&$ 0.5260 \pm 0.0028 $& SB2 EB & 10 \\
YY Gem A \& B     &$ 0.5992 \pm 0.0047 $&$ 0.6191 \pm 0.0057 $& SB2 EB & 11 \\
GU Boo B          &$ 0.599 \pm 0.006   $&$ 0.620 \pm 0.020   $& SB2 EB & 12 \\
GU Boo A          &$ 0.610 \pm 0.007   $&$ 0.623 \pm 0.016   $& SB2 EB & 12 \\

\enddata
\tablerefs{
1. \citet{pont2006}; 2. \citet{pont2005a}; 3. \citet{pont2005b};
4. \citet{lacy1977}; 5. \citet{metcalfe1996}; 6. \citet{bouchy2005};
7. \citet{ribas2003}; 8. \citet{maceroni2004}; 9. \citet{creevey2005};
10. \citet{lopez2006}; 11. \citet{torres2002}; 12. \citet{lopez2005}; 
13. This paper}
\end {deluxetable}

\clearpage

\clearpage

\clearpage

\clearpage

\clearpage

\clearpage

\end{document}